\documentclass[a4paper]{jpconf}

\usepackage{graphicx}

\newcommand{\cp}{\ensuremath{c_{p}}}
\newcommand{\cel}{\ensuremath{c_{\rm el}}}
\newcommand{\cph}{\ensuremath{c_{\rm ph}}}

\newcommand{\gn}{\ensuremath{\gamma_{\rm n}}}
\newcommand{\gs}{\ensuremath{\gamma_{\rm s}}}
\newcommand{\gres}{\ensuremath{\gamma_{\rm res}}}

\newcommand{\TD}{\ensuremath{{\it \Theta}_{\rm D}}}
\newcommand{\Hc}{\ensuremath{H_{\rm c}}}

\newcommand{\Tc}{\ensuremath{T_{\rm c}}}

\newcommand{\HDC}{\ensuremath{H_{\rm DC}}}

\begin{document}

\title{Superconductivity of hexagonal heavily-boron doped silicon carbide}

\author{M~Kriener$^1$, T~Muranaka$^2$, Z-A~Ren$^2$, J~Kato$^2$, J~Akimitsu$^2$ and Y~Maeno$^1$}

\address{$^1$Department of Physics, Graduate School of Science, Kyoto University, Kyoto 606-8502, Japan}
\address{$^2$Department of Physics and Mathematics, Aoyama-Gakuin University, Sagamihara, Kanagawa 229-8558, Japan}

\ead{mkriener@scphys.kyoto-u.ac.jp}

\begin{abstract}
In 2004 the discovery of superconductivity in heavily boron-doped diamond (C:B) led to an increasing interest in the superconducting phases of wide-gap semiconductors. Subsequently superconductivity was found in heavily boron-doped cubic silicon (Si:B) and recently in the stochiometric ''mixture'' of heavily boron-doped silicon carbide (SiC:B). The latter system surprisingly exhibits type-I superconductivity in contrast to the type-II superconductors C:B and Si:B. Here we will focus on the specific heat of two different superconducting samples of boron-doped SiC. One of them contains cubic and hexagonal SiC whereas the other consists mainly of hexagonal SiC without any detectable cubic phase fraction. The electronic specific heat in the superconducting state of both samples SiC:B can be described by either assuming a BCS-type exponentional temperature dependence or a power-law behavior.
\end{abstract}

\section{Introduction}

The three heavily boron-doped semiconductors diamond \cite{ekimov04a}, cubic silicon \cite{bustarret06a}, and silicon carbide \cite{ren07a} belong to the newly discovered family of superconductors based on the diamond structure. The remarkable difference between them is the nature of the superconducting ground state: C:B and Si:B are type-II whereas SiC:B is a type-I superconductor. Silicon carbide itself is a well-known example for polytypism. More than 200 crystal modifications with energetically slightly different ground states are reported in literature. The most common ones are 3C-SiC, 2H-, 4H- and 6H-SiC, and 15R-SiC. The number in front of C (\,=\,cubic unit cell), H (\,=\,hexagonal), and R (\,=\,rhombohedral) indicates the number of Si\,--\,C bilayers stacking in the conventional unit cell. Whereas the cubic structure seems to be a precursor or precondition for the occurrence of superconductivity in the parent systems C:B and Si:B, in SiC:B hexagonal modifications contribute to the superconductivity, too, as we will discuss in this paper. We focus on two different polycrystalline samples. One, referred to as SiC-1, contains three different phase fractions: 3C-SiC, 6H-SiC, and Si and is identical to the sample used in Refs.\,\cite{ren07a} and \cite {kriener08a}, where the preparation details are given. The second sample, referred to as 6H-SiC and prepared in a similar way, is also a multiphase sample mainly consisting of hexagonal 6H-SiC \cite{muranakacomment}. In addition, we identified pure Si and 15R-SiC by x-ray diffraction, but there is no indication of a cubic 3C phase fraction in this sample. In spite of these differences both samples become superconducting at about $\Tc\approx 1.45$\,K and are type-I superconductors as indicated by the observation of a strong supercooling effect in finite magnetic fields in resistivity and AC susceptibility \cite{ren07a,kriener08a,muranakacomment}. The critical field strength was estimated to be about 115\,Oe (SiC-1) and 125\,Oe (6H-SiC) and the residual resistivity $\rho_0$ at \Tc\ amounts to $\sim 0.06$\,m$\Omega$cm (SiC-1) and $\sim 1.2$\,m$\Omega$cm (6H-SiC). The charge-carrier concentration is $1.91\cdot 10^{21}$\,cm$^{-3}$ for SiC-1 and $0.25\cdot 10^{21}$\,cm$^{-3}$ for 6H-SiC as estimated from Hall-effect measurements. The latter value is surprisingly low, only 1/10 of the value measured for SiC-1.

The specific-heat data presented in this paper was taken by a relaxation-time method using a commercial system (Quantum Design, PPMS).

\section{Specific heat}
\begin{figure}
\begin{center}
\includegraphics[width=14.5cm]{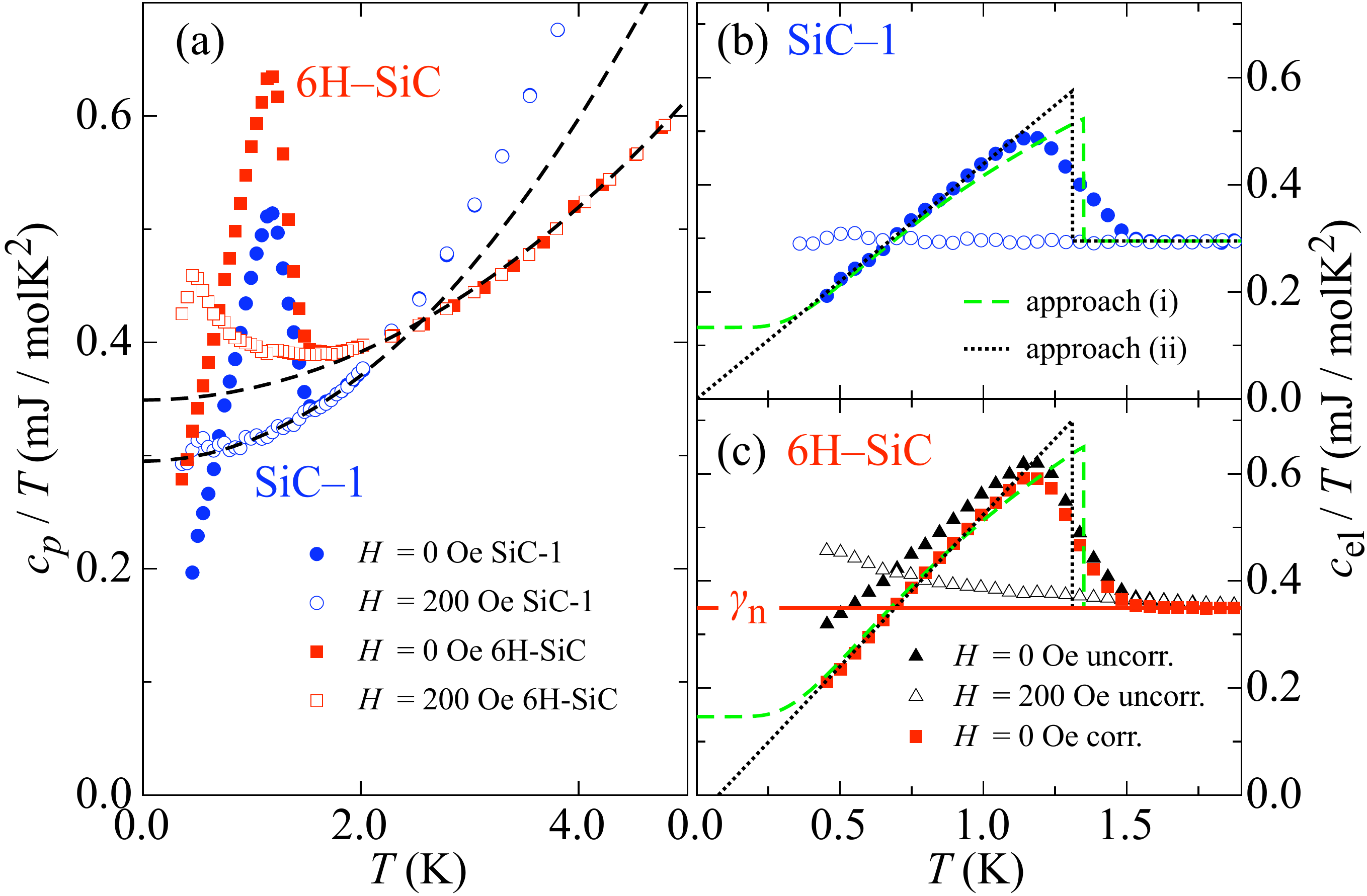}
\end{center}
\caption{\label{fig1} (Color online) Specific heat of the samples SiC-1 and 6H-SiC: The closed symbols in all panels denote data measured in zero field. Open symbols refer to data measured at $H=200\,{\rm Oe}>\Hc$ representing the normal-state specific heat. In panel (a) the specific heat $\cp/T$ as measured is shown. The dashed black curves are results of Debye fits to the data. Panels (b) (SiC-1) and (c) (6H-SiC) contain the electronic specific heat $\cel/T$ and fit results, see text. For 6H-SiC the experimental data (triangles) was corrected (squares) due to experimental problems, see text.}
\end{figure}
In Fig.\,\ref{fig1} specific-heat data of the samples SiC-1 and 6H-SiC are shown. The solid symbols in all panels refer to data taken in zero field, the open symbols denote the normal-state specific heat achieved by applying an external magnetic DC field $\HDC=200$\,Oe\,$>\Hc$. Both samples exhibit a clear jump at \Tc\ as seen in Fig.\,\ref{fig1}\,(a) indicating that the superconductivity in these compounds is a bulk feature. The respective transition in \cp\ is rather broad, reflecting their polycrystalline multiphase character. The in-field data of 6H-SiC exhibits an unusual upwards slope upon decreasing temperature below approximately 2\,K and around 0.4\,K an anomaly occurs. Currently both observations are believed to be the result of experimental problems with our PPMS since we find similar anomalies measuring different samples. 

A fit to the data of SiC-1 in the temperature interval $0.6\,{\rm K} < T < 2$\,K applying the conventional Debye formula $\cp=\cph+\cel = \gn T+\beta T^3$
yields the (normal-state) Sommerfeld parameter $\gn($SiC-1$)=0.29$\,mJ/molK$^2$ and the prefactor of the phononic contribution to the specific heat $\beta($SiC-1$) = 0.02$\,mJ/molK$^4$. For the sample 6H-SiC data at higher temperature $2\,{\rm K} < T < 10$\,K was chosen and a similar Debye fit yields $\gn($6H-SiC$)=0.35$\,mJ/molK$^2$ and $\beta($6H-SiC$) = 0.01$\,mJ/molK$^4$. Both results are displayed as dashed black lines in Fig.\,\ref{fig1}\,(a).
The Debye temperature evaluates to $\TD\approx 590$\,K for SiC-1 and $\approx 715$\,K for 6H-SiC somewhat higher than that found for SiC-1 reflecting the different slope above \Tc. For undoped SiC a Debye temperature of $\TD\approx 1200$\,K\,--\,1300\,K depending on the particular polytype is reported. Therefore the question arises which process is responsible for the strong suppression of \TD\ in this system. For superconducting diamond an earlier specific-heat study \cite{sidorov05a} reports a similar reduction of \TD, whereas a very recent study \cite{dubrovinskaia08a} does not find such a decrease questioning the speculation that a strong suppression of the Debye temperature and hence a strong softening of the corresponding phonon modes is a common effect in this family of superconductors.
Subtracting the phononic contribution from the experimental data yields the electronic specific heat $\cel=\cp-\cph$ displayed in Fig.\,\ref{fig1}\,(b) (SiC-1) and (c) (6H-SiC) as $\cel/T$ vs $T$. 
Due to the mentioned experimental problems in the measurement of 6H-SiC a further analysis of the low-temperature data (black triangles in Fig.\,\ref{fig1}\,(c)) was difficult. Therefore we replaced the in-field electronic specific-heat data (open triangles) by the normal-state Sommerfeld parameter $\gn($6H-SiC$)=0.35$\,mJ/molK$^2$ as obtained from the Debye fit (solid red line in Fig.\,\ref{fig1}\,(c)). Next the difference between the in-field data and \gn(6H-SiC) was calculated and subtracted from the zero-field electronic specific-heat data (closed triangles) assuming that the same background signal is included to the zero-field data. The two data points below the anomaly in the in-field data in Fig.\,\ref{fig1}\,(a) were neglected in this process. This procedure yields the data given in red closed squares in Fig.\,\ref{fig1}\,(c) which will be the base for the analysis carried out next.

An entropy conserving construction (not shown) yields a jump height at \Tc\ of about 1 for both samples, clearly smaller than the BCS weak-coupling expectation 1.43.
We note that for superconducting diamond a jump height of only 0.5 is reported \cite{sidorov05a}. 

For further analyzing the specific-heat data we choose two different approaches as described in detail in Ref.\,\cite{kriener08a}. Approach (i) assumes a BCS-type behavior of the electronic specific heat below \Tc\
\begin{equation}\label{GlBCS_res}
\cel(T)/T = \gres + \gs/\gn\cdot \cel^{\rm BCS}(T)/T.
\end{equation}
The samples used are multiphase samples. Hence it is possible that parts of the samples remain normal conducting allowing for an additional residual contribution to the specific heat below \Tc. To pay respect to this we include an additional residual term $\gres=\gs+\gn$ to Eq.\,\ref{GlBCS_res}. The prefactor \gs\ denotes the contribution of the superconducting parts of the samples. Please note that \gres\ is the only adjustable parameter in this scenario. Approach (ii) assumes a power-law behavior of the electronic specific heat with in principle  three independent fitting parameters
\begin{equation}\label{Glpower}
\cel(T)/T = \gres + a\cdot T^b.
\end{equation}
At low temperatures $b=1$ or 2 corresponds to line or point nodes. For sample SiC-1 both models describe the data reasonably well as can be seen in Fig.\,\ref{fig1}\,(b). The dashed green curve corresponds to approach (i), the dotted black curve to approach (ii). Approach (i) results in a residual contribution $\gres=0.5\times\gn$(SiC-1) corresponding to a superconducting volume fraction of approximately 50\,\%. However, the assumption of a $T$-linear behavior below \Tc\ yields a very good description of the data extrapolating to zero for $T\rightarrow 0$. The fit corresponding to the second approach was done with keeping the exponent $b=1$ in Eq.\,\ref{Glpower}. It is quite surprising that the linear behavior holds up to approximately 1.1\,K, i.\,e.\ up to the transition. In a nodal gap scenario it is expected that for $T\rightarrow \Tc$ the gap magnitude reduces and hence the electronic specific heat deviates from the linear extrapolation. Moreover the fit yields $\gres\approx 0$, i.\,e.\ the fit quality was the same with or without including a residual \gres\ factor emphasizing the surprising strict linear behavior and suggesting a volume fraction of about 100\,\%.

For sample 6H-SiC, as shown in Fig.\,\ref{fig1}\,(c), the fit corresponding to approach (i) yields a reasonable description, too, with a residual contribution of about 40\,\% of the normal-state Sommerfeld parameter. Approach (ii) reveals again a linear $T$ dependence of the electronic specific heat $\cel/T$ in the superconducting state. However, the fit results in a negative value for \gres\ underlining that the rough correction of the data is very speculative. On the other hand the qualitative finding of a power-law behavior somewhat justifies the chosen way. 

We note that an estimation of the superconducting jump height at \Tc\ paying respect to the result of approach (i), i.\,e.\ a finite residual contribution combined with a BCS-like behavior of the specific heat, yields for both specimen approximately 1.48, close to the BCS expectation.

\section{Conclusion}
With the present study we can comment on the question if either the cubic or the hexagonal or even both phase fractions participate in the superconductivity of heavily boron-doped silicon carbide, cf. Ref.~\cite{kriener08a}. Here we demonstrated that hexagonal boron-doped 6H-SiC is a bulk  superconductor as indicated by a clear jump at \Tc. Moreover, approach (ii) of our analysis of the specific heat suggests that the cubic phase fraction in SiC-1 becomes superconducting, too. However, it is possible but would be rather surprising if both phase fractions of SiC-1 exhibit an identical critical temperature since we find only one single sharp transition in our AC susceptibility data of sample SiC-1 \cite{ren07a}. Therefore a comprehensive answer of this question needs further clarification.

In summary, we present a comparative specific-heat study on two different samples of heavily boron-doped SiC. One of them consists of cubic 3C- and hexagonal 6H-SiC whereas the other contains 6H-SiC but no cubic phase fraction. Both exhibit a similar critical temperature and field strength and are type-I superconductors. The electronic specific heat in the superconducting state can be described by either the assumption of a BCS-like exponential temperature dependence including a residual density of states due to non-superconducting parts of the sample or by a power-law behavior.   

\section*{Acknowledgements} 
This work was supported by a Grants-in-Aid for the Global COE ''The Next Generation of Physics, Spun from Universality and Emergence'' from the Ministry of Education, Culture, Sports, Science, and Technology (MEXT) of Japan, and by the 21st century COE program ''High-Tech Research Center'' Project for Private Universities: matching fund subsidy from MEXT. It has also been supported by Grants-in-Aid for Scientific Research from MEXT and from the Japan Society for the Promotion of Science (JSPS). TM is supported by Grant-in-Aid for Young Scientists (B) (No. 20740202) from MEXT and MK is financially supported as a JSPS Postdoctoral Research Fellow.

\section*{References}
\bibliographystyle{/PaperBase/bst/iopart-num}
\bibliography{/PaperBase/preload,/PaperBase/SiC-Si-C,/PaperBase/Superconductivity,/PaperBase/Ruthenate,/PaperBase/sonstigePaper,/PaperBase/Lehrbuch,additionalbib}


\end{document}